\documentclass[border=10mm,11pt, varwidth]{amsart}
\usepackage[left=1in,top=1in,right=1in,bottom=1in]{geometry}
\usepackage{amsmath, amsthm, amssymb, enumitem, bm}
\usepackage[mathscr]{eucal}
\usepackage[square, numbers]{natbib}
\usepackage{float}
\usepackage{placeins} 
\usepackage{tikz}
\usetikzlibrary{arrows.meta}

\usepackage{textcomp}
\usepackage{macro}
\usepackage{url}
\usepackage{braket}
\usepackage{dsfont}
\usepackage[textwidth=1in, textsize=small]{todonotes}
\usepackage{easylist}
\usepackage[compatibility=false]{caption}
\usepackage{subcaption}

\usepackage[linesnumbered,ruled,commentsnumbered,noline]{algorithm2e}

\usepackage{tikz}
\usetikzlibrary{arrows.meta}
\usetikzlibrary{automata, positioning, arrows}
\tikzset{
->, 
node distance=4cm, 
every state/.style={thick, fill=gray!10,scale=1.5}, 
initial text=$ $, 
}

\title[Nonparametric learning of covariate-based Markov jump processes]{Nonparametric learning of covariate-based Markov jump processes using RKHS techniques}

\author{Yuchen Han }
\address{Department of Bioinformatics and Biostatistics, University of Louisville}
\email{yuchen.han@louisville.edu}

\author{Arnab Ganguly}
\address{Department of Mathematics, Louisiana State University, USA.}
\email{aganguly@lsu.edu}
\thanks{Research of A. Ganguly is supported in part by NSF DMS - 2246815 and Simons Foundation (via Travel Support for Mathematicians)}

\author{Riten Mitra}
\address{Department of Bioinformatics and Biostatistics, University of Louisville}
\email{ritendranath.mitra@louisville.edu}

\keywords{Bayesian inference, covariate driven, kernel methods, Markov jump processes, multistate models, nonparametric estimation, RKHS, survival analysis.
}

\begin{document}

\begin{abstract}
We propose a novel nonparametric approach for linking covariates to Continuous Time Markov Chains (CTMCs) using the mathematical framework of Reproducing Kernel Hilbert Spaces (RKHS). CTMCs provide a robust framework for modeling transitions across clinical or behavioral states, but traditional multistate models often rely on linear relationships. In contrast, we use a generalized Representer Theorem to enable tractable inference in functional space. For the Frequentist version, we apply normed square penalties, while for the Bayesian version, we explore sparsity inducing spike and slab priors. Due to the computational challenges posed by high-dimensional spaces, we successfully adapt the Expectation Maximization Variable Selection (EMVS) algorithm to efficiently identify the posterior mode. We demonstrate the effectiveness of our method through extensive simulation studies and an application to follicular cell lymphoma data. Our performance metrics include the normalized difference between estimated and true nonlinear transition functions, as well as the difference in the probability of getting absorbed in one the final states, capturing the ability of our approach to predict long-term behaviors. 
\end{abstract}
\date{\today}

\maketitle
\section{Introduction} \label{sec:intro}
{\em Continuous time Markov Chains (CTMCs)} offer a rigorous mathematical framework for modeling transitions across different clinical or behavioral states. For a two-state Markov Chain, the relationship between the transition rates and relevant covariates has traditionally been handled through the Cox proportional hazards (Cox-PH) model \citep{cox1972regression} with a linear link function. Such modeling has been used extensively in a wide class of time-to-event setups and is central to domains like survival analysis and reliability analysis.

Despite the complexity posed by multiple states and transitions in a general CTMC — particularly in terms of parameter interpretability and predictive probabilities — applications of these models have surfaced in the literature since the 1980s \citep{kalbfleisch1985analysis, marshall1995multi}. In clinical contexts, these models are commonly referred to as {\em multistate models}. Accordingly, we use the terms {\em CTMC} and {\em multistate model} interchangeably in our ensuing discussion. \citep{meira2009multi} provided a comprehensive review that highlighted the critical insights gained by such approaches over traditional two-state survival models. For instance, in the Stanford heart transplant study, the transplant status was modeled as an intermediate state between survival and death. More elaborate applications, such as the European Bone Marrow Transplant (EBMT) registry \citep{fico}, featured richer state configurations involving complete remission, relapse, death, and intermediary recovery stages. The EBMT leukemia dataset also served as an example dataset for the software R package {\em mstate} \citep{de2010mstate,putter2007tutorial}, which remains a standard tool for parametric frequentist multistate estimation. In other applications, the states may represent stages of progressive diseases, as in diabetic retinopathy \citep{marshall1995multi}, or counts of metabolic comorbidities for patients with acute myeloid leukemia \citep{le2018application}. Covariates in such multistate models typically include baseline demographic characteristics, treatment types, and time-varying dose regimens. Recently, \citep{kaur22} demonstrated an interesting application of multistate models in analyzing COVID-19 discharge types, modeling transition rates based on education level and therapist attributes.

Relevant methodologies in literature cover a wide set of  parametric and semi-parametric approaches to estimate transition intensities under both Frequentist and Bayesian paradigm. For example, \citep{kalbfleisch1988likelihood} developed a pseudolikelihood approach which had good efficiency properties compared to fully parametric methods, while \citep{kneib2008bayesian} incorporated penalized splines to estimate baseline transition intensities in a Bayesian setting. 
\citep{crem} proposed a joint semiparametric model for several possibly related multi-state processes by specifying a joint prior distribution on
the transition rates of each process.
 To further investigate how patient characteristics impact the disease progression, a common strategy employed by researchers is to model the covariate effects with linear regression,  as in \citep{aalen2001covariate}, where they estimated transition-specific cumulative regression function for the effect of time varying covariates in a lung cancer study. Also  \citep{hatami2024scalable} leverages the CTMC models coupled with stochastic gradient descent algorithm to fit  clinical prognostic factors in a large scale multiple sclerosis patient data. \citep{zhao2016bayesian} implemented CTMC in phylogenetic setup and conducted Bayesian optimization with Hamiltonian Monte Carlo algorithm. These types of covariate dependence model are log linear in nature and have uniformly marginalized covariate effect per transition. None of these aforementioned researches has reached the individual level of granularity in terms of obtaining individualized transition trajectory by fully utilizing covariate in continuous domain. 

 In real-world clinical settings, the influence of patient characteristics on disease progression often exhibits complex and nonlinear patterns that cannot be adequately captured by simple linear predictors. Conventional approaches are therefore limited in their ability to recover individualized transition dynamics. None of the aforementioned works fully exploit the continuous nature of covariate information to produce individualized, trajectory-level transition rates across states.
 
Motivated by these gaps, and inspired by \citep{ganguly2023infinite},  we propose a novel nonparametric framework based on kernel methods, wherein the transition rate functions $\gen_{i,j}(\cdot)$ are modeled as elements of an appropriate Reproducing Kernel Hilbert Space (RKHS). The structure of RKHS endows the model with favorable analytical properties, and crucially enables the reformulation of the original infinite-dimensional optimization problem into a finite-dimensional one via a generalization of the classical Representer Theorem (RT).  RKHS methods have gained increasing popularity in nonparametric statistics, particularly in nonparametric regression where the goal is to model the mean functional relationship between inputs and outputs via $y=g(x)+\epsilon$ without assuming a specific form for $g$. A novel method of learning Stochastic Differential Equations (SDEs) driven by Brownian motions using RKHS techniques was recently proposed in   \citep{ganguly2023infinite}.  It was shown that the Euler-Maruyama discretization of the SDE, combined with a generalized RT, leads to a closed-form solution for optimizing the objective function in the frequentist framework, as well as an easy-to-implement Gibbs sampling algorithm for Bayesian inference using specific shrinkage priors.

We adapt these RKHS-based tools to the CTMC setup with a finite state space. While the RT can still be successfully applied to the CTMC likelihood (see Theorem \ref{th:rep-est-multi}), the specific structure of the likelihood function precludes a fully closed-form solution to the resulting optimization problem. This introduces significant computational challenges in developing tractable inference schemes for covariate-based Markov processes. 

For frequentist inference, we use the  quasi-Newton gradient descent algorithm to optimize the penalized likelihood.  
For Bayesian inference, we investigate spike-and-slab priors — a specialized class of sparsity-inducing priors — which parallel the role of penalties in the frequentist setting. These priors are especially useful for mitigating the challenges of high-dimensional function estimation. A brief review of such sparsity-inducing and shrinkage priors can be found in \citep{VOM19, ganguly2023infinite}. The complex form of the posterior leads to very low acceptance rates in standard Metropolis–Hastings algorithms,  causing computational inefficiency. To address this, we propose a more robust approach that adapts the Expectation-Maximization Variable Selection (EMVS) algorithm to our multistate setting. EMVS developed in \citep{rovckova2014emvs} for regression problems can enable rapid identification of a promising region of the posterior through an initial EM run. To our knowledge, this is also the first  use of sparsity priors in the context of covariate-based Markov chains.

We conduct an extensive set of simulations to validate our  approach. Our performance metrics include (i)  a normalized difference of the estimated and true transition functions; and (ii) difference between probability distributions (as measured by Kolmogorov distance) induced by these transitions. The latter essentially captures the ability of our methods to correctly predict the long  term  behavior of the Markov Chain. Our generative models for CTMC simulations use  tree-like structures with true  transition functions having polynomial forms. We obtain low error rates on the probability  metrics  with sample size. 

The rest of the paper is structured as follows. In  Section \ref{mt} we  lay out the  mathematical setup for a general CTMC with covariates governing the transitions. Here we specify the precise algebraic expression of the likelihood function.  In  Section \ref{rkhs} we describe the RKHS based approach to maximization of the likelihood function.  
Section \ref{freb} is devoted to Bayesian procedures and the EMVS algorithm for maximizing the posterior probability.  
Section \ref{sim} provides a validation of our scheme against suitable performance metrics, and Section \ref{cs} presents an application of our method to a case study.


\section{Mathematical framework}\label{mt}
Our base model is a  continuous time Markov chain (CTMC)  process, $X=\{X(t):t\geq 0\}$  taking values in  a finite set  $ \statesp=\{1,2, \ldots, |\statesp|\}$ with generator matrix $\genmat=((\gen(i,j)))$.  
 Thus for any function $g: \statesp \rt \R $, the process $\SC{M}^g$ defined by
\begin{align*}
\SC{M}^g(t) = g(X(t))-g(X(0)) -\int_0^t \genmat g(X(s))\ ds
\end{align*}
is a martingale with respect to the filtration $\{\SC{F}_t=\s(X(s):0\leq s\leq t)\}$.  Recall that for a generator matrix $\genmat$, $\gen(i,j)\geq 0$ for $i\neq j$, and $q(i,i) = -\sum_{j\neq i}q(i,j)$, and 
 $\genmat g:\statesp \rt \R$ is defined as 
 $\genmat g(i)=\sum_{j}\gen(i,j)g(j) = \sum_{j\neq i} q(i,j) (g(j)-g(i)).$ 
 
 Notice that the evolution of the CTMC $X$ can be described as follows. Let $\tau_0=0$, and  $\tau_k$ the $k$-th jump-time of the chain $X$ defined as 
$$\tau_k \equiv \inf\{s>\tau_{k-1}: X(s) \neq X(\tau_{k-1})\}.$$
Then conditional on $X(\tau_{k-1})$, the waiting-time for the next jump, $\tau_k-\tau_{k-1}$, and the new jump-state, $X(\tau_k)$, are independent, and 
\begin{align}
\label{eq:MC-dist}
\begin{aligned}
\tau_k-\tau_{k-1}\big|X(\tau_{k-1})=i_{k-1} \sim&\   \mathrm{Exponential}\lf(|\gen(i_{k-1},i_{k-1})|)=\sum_{j\neq i_{k-1}}q(i_{k-1},j)\ri),\\
\PP\lf(X(\tau_k)=i_k| X(\tau_{k-1})=i_{k-1}\ri) =&\ \gen(i_{k-1},i_k)\big/ |\gen(i_{k-1},i_{k-1})|.
\end{aligned}
\end{align}

\np
 {\em Censoring random variable:} The time span of our data is determined by a {\em censoring} random variable $\cenrv$, with density $\cenden$ and distribution $\cendist$. Notice that $\gden_{\cdot| X(\tau_{k-1})=i_{k-1}}$, conditional density of $(\tau_k \wedge \cenrv, X(\tau_k \wedge \cenrv))\big| X(\tau_{k-1})=i_{k-1}$ on $[0,\infty)\times \statesp$ (with respect to product of Lebesgue and counting measure), is given by
\begin{align}\label{eq:cen-mc-den}
\begin{aligned}
\gden_{\cdot|X(\tau_{k-1})=i_{k-1}}(t,i_k)
=&\ \Big(\exp(|\gen(i_{k-1}, i_{k-1})|t)\gen(i_{k-1}, i_k)(1-\cendist(t))\Big)^{I_{\{i_k \neq i_{k-1}\}}} \\ &\ \times  \Big(\exp(|\gen(i_{k-1}, i_{k-1})|t)\cenden(t))\Big)^{I_{\{i_k = i_{k-1}\}}}.
\end{aligned}
\end{align}

\np
{\em Observation Model:} 
We assume that the dynamics of the Markov chain $X$ depends on an $\R^d$-valued covariate $\covar$; that is, given $Z=z$, $X$ has the generator matrix $\genmat\equiv \genmat(z) =((\gen(i,j;z))).$ Our goal is to learn the matrix-valued function $\genmat(\cdot)$, or equivalently, $|\statesp|^2 -|\statesp|$ functions, $\gen(i,j; \cdot),\ i,j=1,2\hdots, |\statesp|, i\neq j$. 

Our data comprises of $L$ trajectories from the Markov chain each of whose dynamics is dictated by a realization of $\covar$, and  each of which is observed until a random time determined by the censoring random variable, $\cenrv$.  Note that a trajectory of the CTMC $X$ can be described by a sequence of pairs, each consisting of jump-times of the chain and the corresponding jump-states. Thus, mathematically, our dataset $\SC{D}$ consists of 
\begin{align*}
\lf\{z^{(l)}, \lf(t^{(l)}_k, i^{(l)}_k\ri): k=1,2,\hdots, K^{(l)}\ri\}, \quad l=1,2,\hdots,L
\end{align*}
$L$ independent realizations of $\lf\{Z, (\tau_k \wedge \cenrv, X(\tau_k \wedge \cenrv)):k=1,2,\hdots\ri\}$, where $X|Z =z^{(l)}$ is a CTMC with generator matrix $\genmat(z^{(l)})$.  \eqref{eq:MC-dist} and \eqref{eq:cen-mc-den} readily give the  likelihood function of $\genmat(\cdot)$,  as
\begin{align}
\label{eq:like}
\begin{aligned}
\lik(\genmat(\cdot)|\SC{D})=&\ \prod_{l=1}^L \prod_{k=1}^{K^{(l)}-1}\exp\lf\{|\gen(i^{(l)}_{k-1}, i^{(l)}_{k-1}; z^{(l)})|\lf(t^{(l)}_k-t^{(l)}_{k-1}\ri)\ri\}\gen(i^{(l)}_{k-1}, i^{(l)}_k; z^{(l)})\\
& \times \Big(\exp\lf\{|\gen(i^{(l)}_{K^{(l)}-1}, i^{(l)}_{K^{(l)}-1}; z^{(l)})|\lf(t^{(l)}_{K^{(l)}}-t^{(l)}_{K^{(l)}-1}\ri)\ri\}\\
& \hs{.7cm} \times \gen(i^{(l)}_{K^{(l)}-1}, i^{(l)}_{K^{(l)}}; z^{(l)})(1-\cendist(t^{(l)}_{K^{(l)}}))\Big)^{I\lf\{i^{(l)}_{K^{(l)}} \neq i^{(l)}_{K^{(l)}-1}\ri\}}\\
& \times \Big(\exp\lf\{|\gen(i^{(l)}_{K^{(l)}-1}, i^{(l)}_{K^{(l)}-1}; z^{(l)})|\lf(t^{(l)}_{K^{(l)}}-t^{(l)}_{K^{(l)}-1}\ri)\ri\}\cenden(t^{(l)}_{K^{(l)}})\Big)^{I\lf\{i^{(l)}_{K^{(l)}} = i^{(l)}_{K^{(l)}-1}\ri\}}.
\end{aligned}
\end{align}

\section{Minimization in RKHS} \label{rkhs}
Our objective is to uncover the functional dependence of the rates  $\gen(i,j; \cdot),\ i\neq j$  on the covariate $Z$. Learning of the functions $\gen(i,j; \cdot): \R^d \rt [0,\infty), \ i\neq j$ requires setting up a suitable infinite-dimensional minimization problem over Hilbert space-valued functions.
To ensure non-negativity of the $\gen(i,j,\cdot)$ for $i\neq j$, we set the $\genmat$-matrix as 
\begin{align}
\label{eq:lngen-def}
\gen(i,j,\cdot) = 
\begin{cases}
\exp\lf(\lngen(i,j;\cdot)\ri),& \quad i \neq j,\\
-\sum_{i'\neq i} \gen(i,i',\cdot),& \quad i = j,
\end{cases}
\end{align}
 where $\lngen(i,j;\ \cdot): \R^d \rt \R$. To write this in a vector-notation for convenience, we introduce the function $\EXP:\R^{|\statesp|^2-|\statesp|} \rt \R^{|\statesp|\times |\statesp|}$ and write $\genmat(\cdot) = \EXP(\bm{\lngen}(\cdot))$ to represent \eqref{eq:lngen-def}, where  $\bm{\lngen}(\cdot): \R^d \rt \R^{|\statesp|^2-|\statesp|}$, defined by $ \bm{\lngen}(\cdot) =\ve\lf(\lngen(i,j,\cdot), (i,j) \in \statesp\times \statesp, i \neq j\ri),$
  denotes the vectorization of the collection$\{\lngen(i,j;\cdot)=\ln \gen(i,j,\cdot), i\neq j\}.$

  Our approach to learning the vector-valued function $\bm{\lngen}$ relies on optimization on a reproducing kernel Hilbert space (RKHS) of vector-valued functions. The RKHS of vector-valued functions generalizes the notion of RKHS of real-valued functions, with the key distinction that the associated kernel is matrix-valued.

\begin{definition} \label{def:matRK}
	Let $\meU$ be an arbitrary input space.
	A symmetric function $\bm{\knl}: \meU\times \meU \rt \R^{n\times n}$ is a {\em reproducing kernel} if for any $u, u' \in \meU$, $\bm{\knl}(u,u')$ is a $n\times n$ p.s.d. matrix.
	
	The RKHS associated with a reproducing kernel $\bm{\knl}$ is a Hilbert space $\Hsp_{\bm{\knl}}$ of functions $h: \meU \rt \R^n$, such that  for every fixed $u \in \meU$ and a (column) vector $v \in \R^n$, (i) the mapping $u' \rt \bm{\knl}(u',u)v$ is an element of $\Hsp_{\bm{\knl}}$, and (ii) $\<h, \bm{\knl}(\cdot, u)v\> = h(u)^Tv.$
\end{definition}
Property (ii) corresponds to the reproducing property of the kernel $\bm{\knl}$ within the vector-valued framework. Similar to the scalar case, a reproducing matrix-valued kernel, $\bm{\knl}$, allows the construction of the associated RKHS via an extension of the Moore-Aronszajn theorem. The RKHS is given by $\Hsp_{\bm{\knl}} = \overline{\text{Span}}\{\bm{\knl}(\cdot,u): u \in \meU\}$, where the closure is taken with respect to the norm, $\|\cdot\|_{\bm{\knl}}$, defined
$$\|h\|_{\bm{\knl}} = \sum_{i,j=1}^l c_i^T \bm{\knl}(u_i,u_j)c_j, \quad h = \sum_{j=1}^l \bm{\knl}(\cdot, u_j)c_j, \ c_j \in \R^n.$$

The following theorem, which is a direct consequence of \citep[Corollary 12]{ganguly2023infinite} --- a generalized RT for RKHS of vector-valued functions, is the center piece of our algorithm.

\begin{theorem} \label{th:rep-est-multi}
Let $\bm{\knl}: \R^d\times \R^d \rt \R^{(|\statesp|^2-|\statesp|)\times (|\statesp|^2-|\statesp|)}$ be a matrix-valued reproducing kernel with associated RKHS of $\R^{(|\statesp|^2-|\statesp|)\times (|\statesp|^2-|\statesp|)}$-valued functions, $\Hsp_{\bm{\knl}}$. Define the loss function $\loss: \R^{(|\statesp|^2-|\statesp|)\times (|\statesp|^2-|\statesp|)} \rt \R$ by
\begin{align}
\label{eq:loss}
\loss(\bm{\lngen}(\cdot)|\SC{D}) = - \ln \lik\Big(\genmat(\cdot)=\EXP(\bm{\lngen}(\cdot))\Big|\SC{D}\Big).
\end{align}

Consider the optimization problem
\begin{align}
\label{eq:main-opt}
\argmin_{\bm{\lngen} \in \Hsp_{\bm{\knl}} } \loss(\bm{\lngen}(\cdot)|\SC{D}).
\end{align}
Then $\bm{\lngen}$ admits the finite-sum representation
\begin{align}
\label{eq:fin-rep-min}
\bm{\lngen}(z) = \sum_{l=1}^L \bm{\knl}(z,z^{(l)})\wtb_l,
\end{align}
where the coefficients $\wtb^{(l)} \in \R^{(|\statesp|^2-|\statesp|)}$.
\end{theorem}

In practice, the $(|\statesp|^2-|\statesp|)\times (|\statesp|^2-|\statesp|)$-matrix valued kernel $\bm{\knl}$ is taken to be $\bm{\knl}=\mathrm{diag}(\knl_{(i,j)}: i, j = 1,2,\hdots, |\statesp|, i\neq j )$ in which case \eqref{eq:fin-rep-min} takes the following simpler form:  for each $i\neq j$,
\begin{align}
\label{eq:fin-rep-min-2}
\lngen(i,j; z) = \sum_{l=1}^L \knl_{(i,j)}(z,z^{(l)})\wtb^{(i,j)}_l, \quad \wtb^{(i,j)}_l \in \R,
\end{align}
with $\wtb_l$ being the $|\statesp|^2-|\statesp|$-dimensional vector with entries $\wtb^{(i,j)}_l, i, j = 1,2,\hdots, |\statesp|, i\neq j.$ Let $\bigpara$ be an $L(|\statesp|^2-|\statesp|)$-dimensional vector created by stacking up $(|\statesp|^2-|\statesp|)$-dimensional $\wtb_l, \ l=1,2,\hdots, L$, that is,
$$\bigpara^T \dfeq  (\wtb^T_1,\wtb^T_2, \hdots,\wtb^T_L).$$
The remarkable feature of Theorem \ref{th:rep-est-multi} is that it transforms the infinite-dimensional optimization  problem, \eqref{eq:main-opt}, to a finite dimensional one,  where learning the function $\bm{\lngen}(\cdot): \R^d \to \R^{|\statesp|^2 - |\statesp|}$ requires only the estimation of the parameter vector $\bigpara$. By a slight abuse of notation, we denote
$$  \loss(\bigpara|\SC{D}) \equiv \loss\lf(\sum_{l=1}^L \bm{\knl}(\cdot,z^{(l)})\wtb_l|\SC{D}\ri),$$
and our goal is to minimize this over $\bigpara \in \R^{L(|\statesp|^2-|\statesp|)}$.

\section{Bayesian approach to estimation} \label{freb}
The frequentist approach to estimation of the parameter vector $\bigpara$ involves minimization of 
$ \loss(\bigpara|\SC{D}) $ via standard optimization techniques like  quasi-Newton gradient descent algorithm like Nelder-Mead or  Broyden–Fletcher–Goldfarb–Shanno (BFGS). This provides a point estimate of $\hat{\bigpara}$. Now sparse estimation of  $\hat{\bigpara}$ is  desirable because it improves interpretability, reduces overfitting, enhances computational efficiency, and focuses computational efforts on the most relevant $\wtb_l$. While sparse solutions can be achieved in the frequentist framework by adding a penalty term (such as $\ell_1$ regularization) to the objective function, such approaches depend on tuning parameters (e.g., the penalty weight) and typically yield a single ``best" model without a natural quantification of uncertainty in the selection.


This paper primarily adopts a Bayesian approach, which naturally addresses the limitations of the frequentist methods discussed above. Building on the expansion \eqref{eq:fin-rep-min}, justified as an optimal representation of the estimator of $\bm{\lngen}$ by Theorem \ref{th:rep-est-multi}, the Bayesian framework involves placing a prior density $\prior$ on the weight vector $\bigpara$ and computing the corresponding posterior density $\post(\bigpara|\SC{D})$, given by
$$\post(\bigpara|\SC{D})=\lik\lf(\genmat(\cdot)=\EXP\lf(\sum_{l=1}^L \bm{\knl}(\cdot,z^{(l)})\wtb_l\ri)\Big|\SC{D}\ri)\prior(\bigpara),$$
where the likelihood functional $\lik$ was defined in  \eqref{eq:like}.

{\em Choice of prior:} In Bayesian analysis, priors serve various purposes, such as incorporating insights from earlier studies, emphasizing certain variables through preferential weighting, or remaining uninformative to facilitate purely data-driven inference.  Inference with uninformative priors often aligns closely with Frequentist approaches. 

In our approach we utilize a special class of structural priors known as variable selection priors that encourage sparse solutions, i.e. many parameters are effectively set to zero. This is particularly necessary for high-dimensional parameter estimation problems like ours where the dimensionality of the parameter vector $\bigpara$ grows with the size of the data set. The purpose of such sparsity-inducing priors is to encourage the estimation procedure  to favor values close to zero for most $\wtb^{(i,j)}_l$. A key characteristic of these priors is their combination of a sharp peak near zero and heavy tails.  The idea behind choosing a distribution with such feature is that  the sharp peak  will be able to shrink the unnecessary coefficients, while the heavy tail pulls the  strong signals  away from  $0$  and thereby uncovering the strong signals.

Within the large class of variable selection priors, shrinkage and spasity-inducing priors form important subcategories. Many shrinkage  priors belong to the   normal scale-mixture family that includes the  $t$-prior \citep{tipp}, double-exponential \citep{park2008bayesian} and Horseshoe priors \citep{carvalho2009handling, carvalho2010horseshoe}. Although shrinkage priors are effective at reducing certain norms of the parameter vector, they typically do not enforce exact zeros in the parameter values -- a property desirable for model interpretability and selection in our setting. Moreover, while shrinkage priors in the class of normal-scale mixtures for regression problems (and also for SDE estimation) allows sampling of posterior distribution through an easy-to-implement Gibbs sampling, this approach is not feasible here as the conditional distribution of the parameters are not available in closed form due to the complexity of the likelihood function $\lik(\cdot|\SC{D})$. To address this, we employ the popular spike and slab prior on $\bigpara,$ which is particularly well-suited to our needs.

The Spike and Slab  prior is   a  mixture of  two Gaussian distributions, namely the  (i) `spike'  $N(0, \nu_0)$-distribution having a low variance $\nu_0$ around 0,  and  (ii) the `slab' $N(0, \nu_1)$- representing a  more diffuse distribution which assigns relatively high probability mass to larger values. Although the variance parameter $\nu_0$ of the spike distribution is often chosen as zero in practice, following \cite{GM97}, we consider a small but positive $\nu_0 > 0$ to enhance the exclusion of insignificant nonzero effects. Using a latent-variable $\gamma$, the Bayesian hierarchy of the prior can be represented as follows:
\begin{align}
\label{eq:prior-specs}
\begin{aligned}
\prior(\wtb^{(i,j}_l)\ =&\ \lf(1-\gamma^{(i,j)}_l\ri)\No\lf(\wtb^{(i,j)}_l|0,\nu_0\ri)+\gamma^{(i,j)}_l\No\lf(\wtb^{(i,j)}_l|0,\nu_1\ri)\\
\gamma^{(i,j)}_l \ \stackrel{i.i.d}\sim&\ \mathrm{Bernoulli}(\berprob)\\
\berprob\ \sim&\ \mathrm{Beta}(a_0,a_1), \quad a_1, a_0 >0.
\end{aligned}
\end{align}
We denote
$$\bm{\gamma} =\ve(\gamma^{(i,j)}_l, i,j=1,2,\hdots, |\statesp|, i\neq j, l=1,2,\hdots,L ), \quad \|\bm{\gamma}\|=\ \sum_{\substack{i,j,l\\ i\neq j}}\gamma^{(i,j)}_l.$$
In our setting, direct estimation of the full posterior density $\post(\bigpara, \berprob, \bm{\gamma}|\SC{D})$, given by
\begin{align}
\label{eq:post-form-full}
\begin{aligned}
\post(\bigpara, \berprob, \bm{\gamma}|\SC{D}) \propto &\  \lik\lf(\genmat(\cdot)=\exp\lf(\sum_{l=1}^L \bm{\knl}(\cdot,z^{(l)})\wtb_l\ri)\Big|\SC{D}\ri)\\
&\times \prod_{\substack{i,j,l\\ i\neq j}}\lf(2\pi( (1-\gamma^{(i,j)}_l )\nu_0+ \gamma^{(i,j)}_l \nu_1))\ri)^{-1/2}\exp\lf(-(\wtb^{(i,j)}_l)^2\big/2( (1-\gamma^{(i,j)}_l )\nu_0+ \gamma^{(i,j)}_l \nu_1))\ri)\\
&\ \times  \berprob^{\|\bm{\gamma}\|}(1-\berprob)^{L(|\statesp|^2-|\statesp|)-\|\bm{\gamma}\|} \times 
 \berprob^{a_0-1}(1-\berprob)^{a_1-1},
\end{aligned}
\end{align}
via traditional MCMC methods is computationally expensive and often inefficient. The high dimensionality of the parameter space and the convoluted form of the likelihood function impair the mixing properties of standard samplers, resulting in slow convergence and poor posterior exploration. Moreover, when the primary inferential goal is variable selection and sparse estimation rather than full posterior uncertainty quantification, exhaustive sampling becomes unnecessarily burdensome.

To overcome these challenges, we adopt a version of the EM-based variable selection (EMVS) algorithm, introduced by Ro{\v{c}}ková and George \citep{rovckova2014emvs} in regression setting,  and extend it to our model. Rather than sampling from the full posterior $\post(\bigpara, \berprob, \bm{\gamma}|\SC{D})$, EMVS reframes posterior exploration as a mode-finding problem with the goal of maximizing the log-posterior $\ln \post(\bigpara, \berprob|\SC{D})$, given by 
\begin{align}
\label{eq:post-form-marg}
\post(\bigpara, \berprob|\SC{D}) =\sum_{ \substack{\gamma^{(i,j)}_l \in \{0,1\}\\ (i,j,l), i\neq j} } \post(\bigpara, \berprob, \bm{\gamma}|\SC{D}).
\end{align}
 In this formulation, the vector of latent variables $\bm{\gamma}$ is treated as missing, and $\post(\bigpara, \berprob|\SC{D})$ in \eqref{eq:post-form-marg} represents the joint posterior of $\bigpara$ and the hyperparameter $\berprob$ with $\bm{\gamma}$ being summed over. This sets the stage for the application of the popular EM algorithm  for maximizing $\ln\post(\bigpara, \berprob|\SC{D})$. The iterative method proceeds as follows: given the $k$-th iterate $(\bigpara^{(k)}, \berprob^{(k)})$, the $(k+1)$-th iterate of the estimator of $(\bigpara, \berprob)$, $(\bigpara^{(k+1)}, \berprob^{(k+1)})$, is obtained by maximizing the value function 
$$\val(\bigpara, \berprob| \bigpara^{(k)}, \berprob^{(k)}) \dfeq \EE_{\bm{\gamma} |\cdot}\lf[\ln \post(\bigpara, \berprob, \bm{\gamma}|\SC{D})\ri],$$
where $ \EE_{\bm{\gamma} |\cdot} \equiv  \EE_{\bm{\gamma} |\bigpara^{(k)}, \berprob^{(k)}}$ represents the conditional expectation operator of $\bm{\gamma}$ given the $k$-th iterate $(\bigpara^{(k)}, \berprob^{(k)})$. Notably, EMVS bypasses the need for extensive sampling and directly targets sparse, high-probability regions of the posterior. This not only leads to dramatic computational gains but also improves the stability and robustness of variable selection, particularly in high-dimensional settings where MCMC approaches tend to struggle with multimodality, low acceptance rate and slow mixing.

\np
{\bf E-Step}: The value function can be written as 
\begin{align*}
\val(\bigpara, \berprob| \bigpara^{(k)}, \berprob^{(k)}) =C(\bm{\gamma})+ \val_0(\bigpara, \berprob| \bigpara^{(k)}, \berprob^{(k)}),  
\end{align*}
where $C(\bm{\gamma})$ is a constant not depending on $(\bigpara, \berprob)$ and $\val_0$ is given by
\begin{align*}
  \val_0(\bigpara, \berprob| \bigpara^{(k)}, \berprob^{(k)}) =&\ \ln \lik\lf(\genmat(\cdot)=\exp\lf(\sum_{l=1}^L \bm{\knl}(\cdot,z^{(l)})\wtb_l\ri)\Big|\SC{D}\ri)\\
  &\ -\f{1}{2}\sum_{\substack{i,j,l\\ i\neq j}}(\wtb^{(i,j)}_l)^2 \EE_{\bm{\gamma} |\cdot}\lf(\f{1}{(1-\gamma^{(i,j)}_l )\nu_0+ \gamma^{(i,j)}_l \nu_1}\ri)+ \lf(\EE_{\bm{\gamma} |\cdot}(\|\bm{\gamma}\|)+a_0-1\ri)\ln \berprob\\
  &\ +\lf(L(|\statesp|^2-|\statesp|)+a_1-1-\EE_{\bm{\gamma} |\cdot}(\|\bm{\gamma}\|)\ri)\ln (1-\berprob).
\end{align*}
Notice that
\begin{align*}
\EE_{\bm{\gamma} |\cdot}(\gamma^{(i,j)}_l) =&\  \PP(\gamma^{(i,j)}_l=1|\bigpara^{(k)}, \berprob^{(k)})\\
=&\  \f{\post(\wtb^{(k,i,j)}_l|\gamma^{(i,j)}_l=1)\PP(\gamma^{(i,j)}_l=1|\berprob^{(k)})}{\post(\wtb^{(k,i,j)}_l|\gamma^{(i,j)}_l=1)\PP(\gamma^{(i,j)}_l=1|\berprob^{(k)})+\post(\wtb^{(k,i,j)}_l|\gamma^{(i,j)}_l=0)\PP(\gamma^{(i,j)}_l=0|\berprob^{(k)})},
\end{align*}
and because $\gamma^{(i,j)}_l$ are binary random variables,
\begin{align*}
 \EE_{\bm{\gamma} |\cdot}\lf(\f{1}{(1-\gamma^{(i,j)}_l )\nu_0+ \gamma^{(i,j)}_l \nu_1}\ri) = \f{\EE_{\bm{\gamma} |\cdot}(1-\gamma^{(i,j)}_l)}{\nu_0}+ \f{\EE_{\bm{\gamma} |\cdot}\gamma^{(i,j)}_l}{\nu_1}.   
\end{align*}

\np
{\bf M-Step:} Clearly,
$(\bigpara^{(k+1)}, \berprob^{(k+1)}) = \argmax \val(\bigpara, \berprob| \bigpara^{(k)}, \berprob^{(k)}) = \argmax \val_0(\bigpara, \berprob| \bigpara^{(k)}, \berprob^{(k)}).$
This maximization problem does not entail a closed form solution and we resort to numerical optimization of $\val_0(\cdot,\cdot|\bigpara^{(k)}, \berprob^{(k)}).$ The literature of EM algorithms applied to  posterior distributions provides some theoretical guarantees of convergence to the posterior mode. However, even in the simpler case  of  linear multivariate regression, posterior inference could be challenging. \citep{rovckova2014emvs}  recommends choosing  optimal  $\nu_0$ and $\nu_1$  by running a grid search over  a range of plausible values.  As an alternative strategy, we set the hyper-parameters  from an initialization procedure. Specifically we run a mixture algorithm on the $L(|\statesp|^2-|\statesp|)$-dimensional $\bigpara$ vector obtained from the  frequentist optimization procedure. Then we apply the mixture algorithm from \textit{mixtools}  package on optimized $\beta$ (treating it as as an univariate data) to obtain two clusters. We  fix $\berprob,\nu_0$, and $\nu_1$ to our cluster means and variances respectively and obtained initialization parameter estimates from running a mixture algorithm. 

\section{SIMULATION STUDY} \label{sim}
\begin{figure}[H] 
\centering
\caption{(a)Basic setting: Three states, Two transitions. Three states corresponds to initial state (1) and two absorbing states (2,3) which are mutually exclusive to each other); (b) Four states, Three transitions. State (2) is intermediary state and states (3,4) are absorbing states.}   \label{fig1} 
\begin{tabular}{c@{\hskip 0.5in}c}
(a)  & (b)  \\
\resizebox{0.3\textwidth}{!}{%
\begin{tikzpicture}[every text node part/.style={align=center}]
\node[state] (1) { $2$};
\node[state, below left of=1] (2) { $1$};
\node[state, right of=2] (3) {$3$};
\draw [-{Stealth[ width=5mm]}]
(2) edge[left] node{$q_{12}(z)$} (1)

(2) edge[below] node{$q_{13}(z)$} (3);
\end{tikzpicture}
}&  
\resizebox{0.5\textwidth}{!}{%
\begin{tikzpicture}[every text node part/.style={align=center}]
\node[state] (1) { $2$};
\node[state, below left of=1] (2) { $1$};
\node[state, right of=2] (3) {$3$};
\node[state, right of=1] (4) {$4$};
\draw [-{Stealth[ width=5mm]}]
(2) edge[left] node{$q_{12}(z)$} (1)

(2) edge[below] node{$q_{13}(z)$} (3)
(1) edge[below] node{$q_{24}(z)$} (4);
\end{tikzpicture}
 }\\
\end{tabular}
\end{figure}
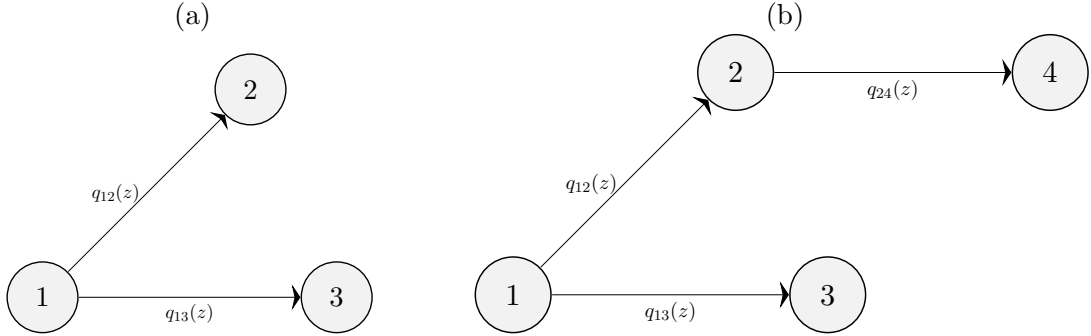

We perform a series of simulation studies to validate our proposed method. Specifically, we consider two settings: a three-state CTMC and a four-state CTMC with a tree structure (see Figure \ref{fig1}).
For the three-state CTMC, we examine two scenarios:
(a) only the $(1,2)$ transition rate function, $\gen_{12}(\cdot)$, is modeled nonparametrically, while the $(1,3)$ transition function, $\gen_{13}$, is assumed known;
(b) both the $(1,2)$ and $(1,3)$ transition functions are treated nonparametrically. In addition, we include a four-state CTMC example, with all three transition arms modeled nonparametrically.  For each scenario, we consider three cases where the true underlying transition functions for the nonparametric arms are quadratic, cubic, and quartic polynomials, respectively. The data is generated by using the Gillespie algorithm with the transition functions listed in Table \ref{tab1}. 

 \begin{table}[H] \centering 
\renewcommand\thetable{1} 
  \caption{Functional forms of transition rates} 
  \label{tab1} 
  \subcaption{$(1,2)$-transition arm nonparametric} 
\begin{tabular}{@{\extracolsep{5pt}}lccc} 
\\[-1.8ex]\hline 
\hline \\[-1.8ex] 
\multicolumn{1}{c}{\small } 
&  \multicolumn{1}{c}{\small $\bm{\lngen}_{12}(z)$}
& \multicolumn{1}{c}{\small $\bm{\lngen}_{13}(z)$}
\\
\hline \\[-1.8ex] 
\small Case 1 & \small$.5+.01z-2z^2$ & \small $.07+.6z$
\\
\hline
\small Case 2 & \small $-2+.1z-.05z^2+z^3$ & \small $.07+.6z$
\\
\hline
\small Case 3 & \small $2+.02z-4z^2+.5z^4$  & \small $.07+.6z$
\\
\hline \\[-1.8ex] 
\end{tabular} 
\vspace{2em}
\subcaption{Three and four state CTMC with all the arms nonparametric} 
\begin{tabular}{@{\extracolsep{5pt}}lccc} 
\\[-1.8ex]\hline 
\hline \\[-1.8ex] 
\multicolumn{1}{c}{\small } 
&  \multicolumn{1}{c}{\small $\bm{\lngen}_{12}(z) =\bm{\lngen}_{13}(z)=\bm{\lngen}_{24}(z)$}
\\
\hline \\[-1.8ex] 
\small Case 1 & \small$.5+.01z-2z^2$ 
\\
\hline
\small Case 2 & \small $-2+.1z-.05z^2+z^3$ 
\\
\hline
\small Case 3 & \small $2+.02z-4z^2+.5z^4$ 
\\
\hline \\[-1.8ex] 
\end{tabular} 
\end{table}

We  fix  $\berprob=.2$ and $\nu_0=.4, \nu_1=5$ to our cluster means and variances respectively and obtained initialization parameter estimates from running a mixture algorithm. We specify the hyper-parameters $a_0=3, a_1=.5$ in \ref{eq:prior-specs} and choose Gaussian or radial basis function (RBF)-kernel for our simulation: $\knl_{(1,2)}(z,z') = \knl_{(1,3)}(z,z') = \exp\lf(-\f{1}{2}(z-z')^2\ri)$.

We used 2 metrics to assess the model performance: mean squared error (MSE) of $\lngen$-functions and and $d_{\mrm{absorption}}$. The metric, $d_{\mrm{absorption}}$, looks at the difference between probabilities of getting absorbed at an absorbing state for the original Markov chain and the Markov chain with estimated transition functions. Specifically, for an absorbing state $a$, $d_{\mrm{absorption}}(a) = |\PP( X(t=\infty) =a) - \PP(\hat X(t=\infty) =a)|$, where $\hat X$ represents the CTMC with estimated $\hat \gen_{ij}(\cdot)$. These probabilities are computed via Monte Carlo approximations based on simulated trajectories from both the true and estimated models, capturing the long-term predictive performance of the estimator. We used 1,000 simulated trajectories for the Monte Carlo estimation of these probabilities.

\begin{figure}[H]
\caption{Plots of the the true $\gen_{12}(\cdot)$ in quadratic, cubic and quartic forms  (black) and the corresponding estimated  version using Bayesian approach ((posterior mode, red) and frequentist approach (blue).} \label{fig2}
\includegraphics[scale=.8]{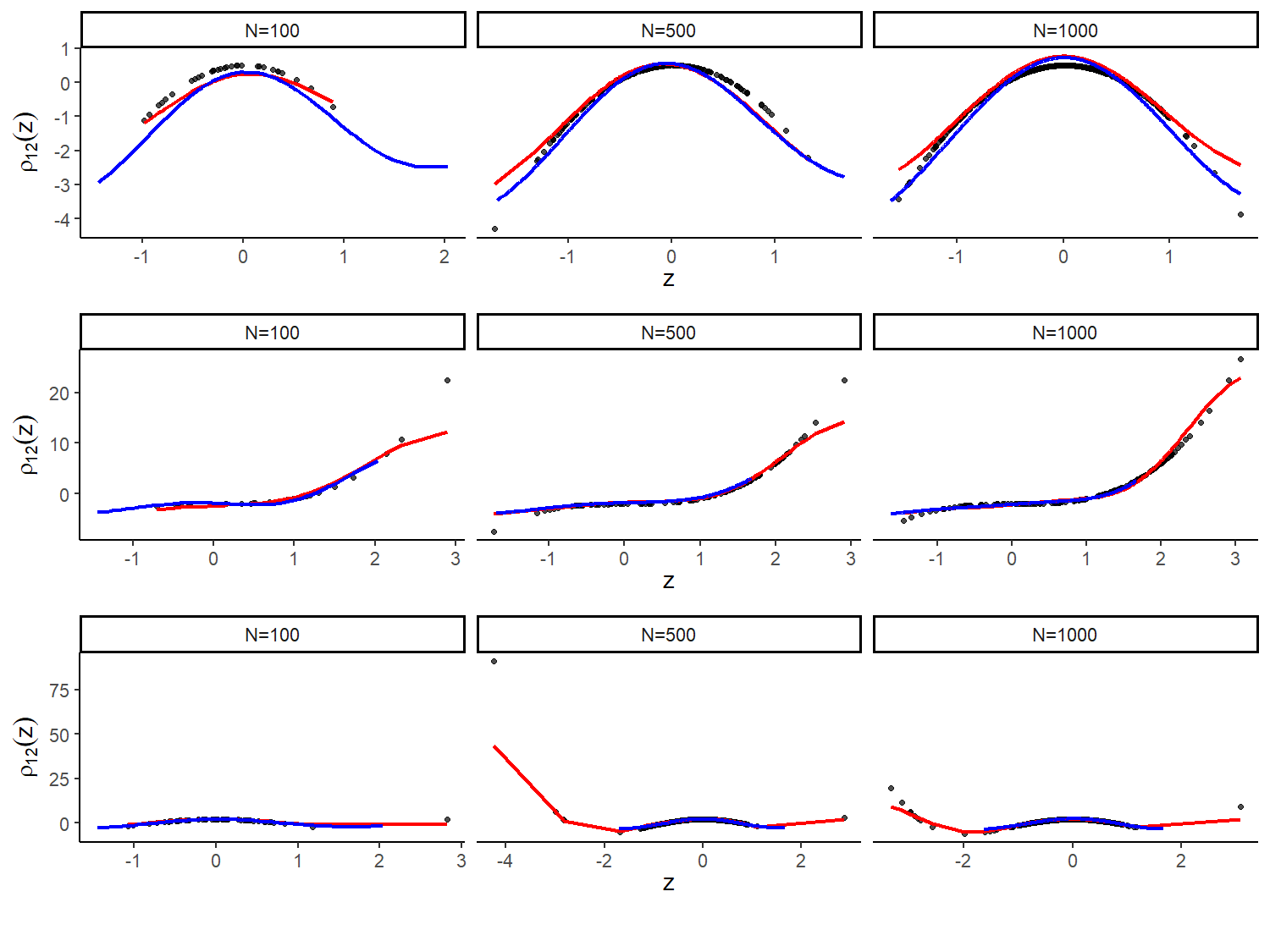}
\end{figure}

A number of visual and numerical  summaries attest strongly to the good performance  of our method.  Figure \ref{fig2}  compares  the true curve of the nonparametric transition arm, $\lngen_{12}(\cdot)=\ln\gen_{12}(\cdot)$, (black) with the estimated ones in Bayesian (red) and Frequentist (blue) settings respectively.  Evidently they are  very similar in regions of high data concentration and tapers off slightly on the extremities with  low data frequencies.

\begin{figure}[H]
\centering
\caption{Plots of averaged MSE for estimated $\lngen_{12}(\cdot)$ over 5 simulated datasets, with true function given by quadratic, cubic and quartic functions in polynomial settings versus sample size N} \label{fig3}
\includegraphics[scale=.8]{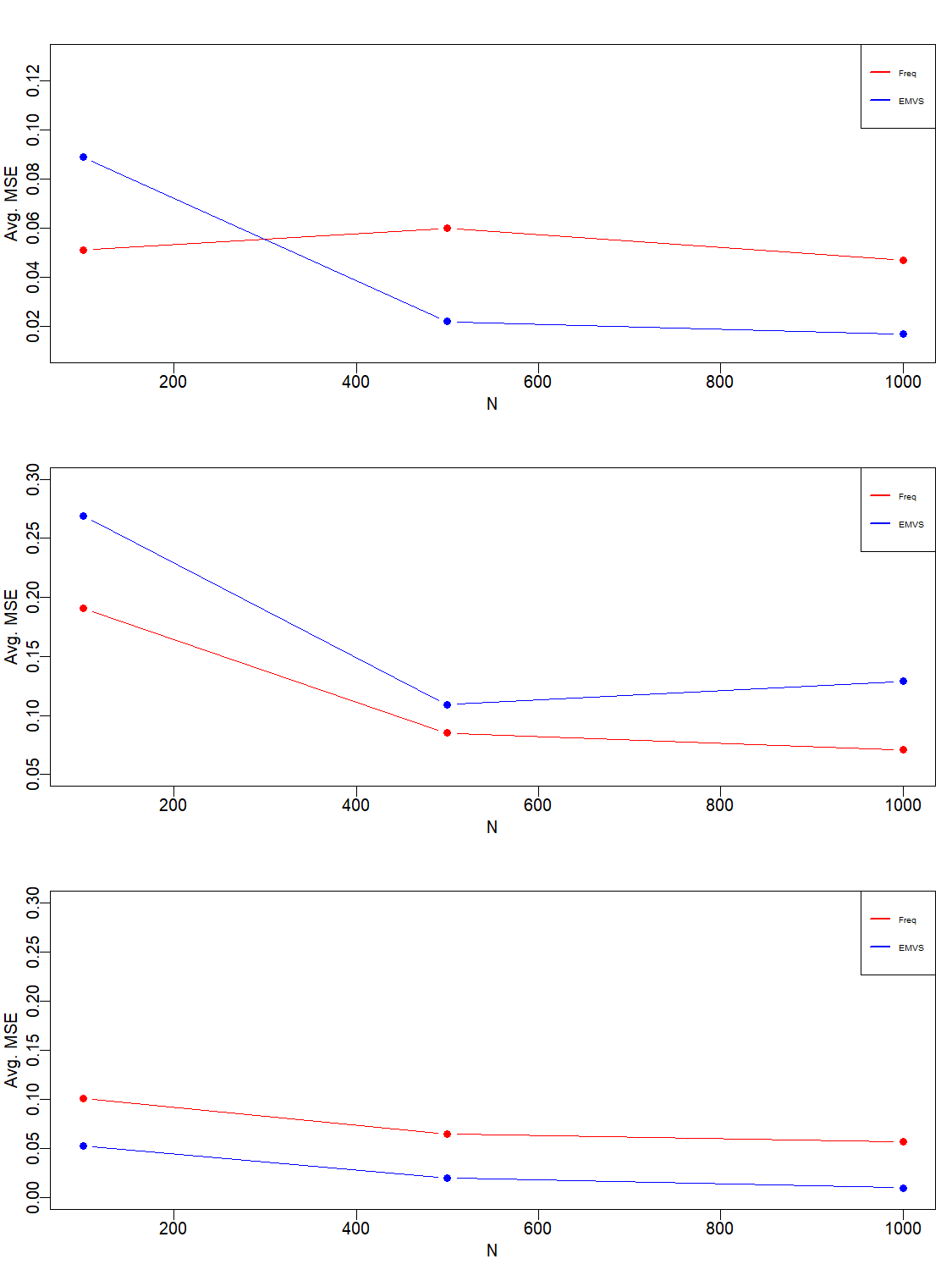}
\end{figure}

Figure \ref{fig3} shows the relationship of the averaged MSE (across 5 replicates) in the three different polynomial settings with one arm treated nonparametric. There is a clear inverse relationship between sample size and error rates when $\lngen_{12}$ is quadratic. This decrease in error rates is markedly  prominent from N=100 to $\sim <600$ and Frequentist estimation performs slightly better than Bayesian  when N is 100. However, it is outperformed  by Bayesian for greater sample sizes (between 200 and 400). While in the cubic setting, the Frequentist estimation generates lower error rates than Bayesian in all sample sizes, demonstrated as the opposite trend in quartic setting.  The results on the both arms reinforce those findings. 

\begin{figure}[H]
\caption{Plots of the the true $\lngen_{12}(\cdot)$ and  $\lngen_{13}(\cdot)$ in quadratic, cubic and quartic forms  (black) and the corresponding estimated  versions using Bayesian approach ((posterior mode, red) and frequentist approach (blue)} \label{fig-2arm}
\includegraphics[scale=.8]{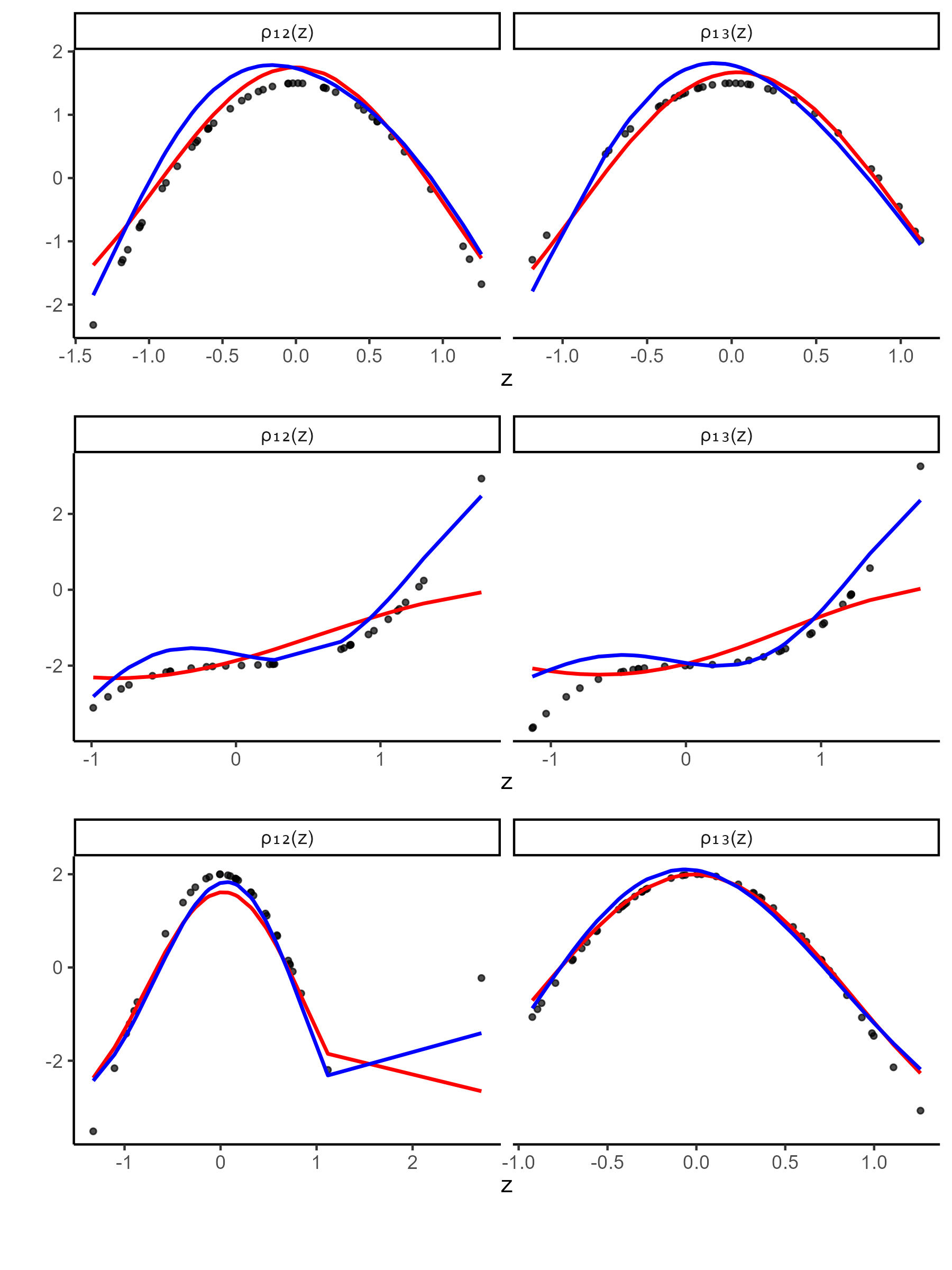}
\end{figure}

\begin{figure}[H]
\centering
\caption{Plots of averaged MSE  for both the estimated arms over 5 simulated datasets, with true function given by quadratic, cubic and quartic functions in polynomial settings versus sample size N} \label{fig-2arm-mse}
\includegraphics[scale=.8]{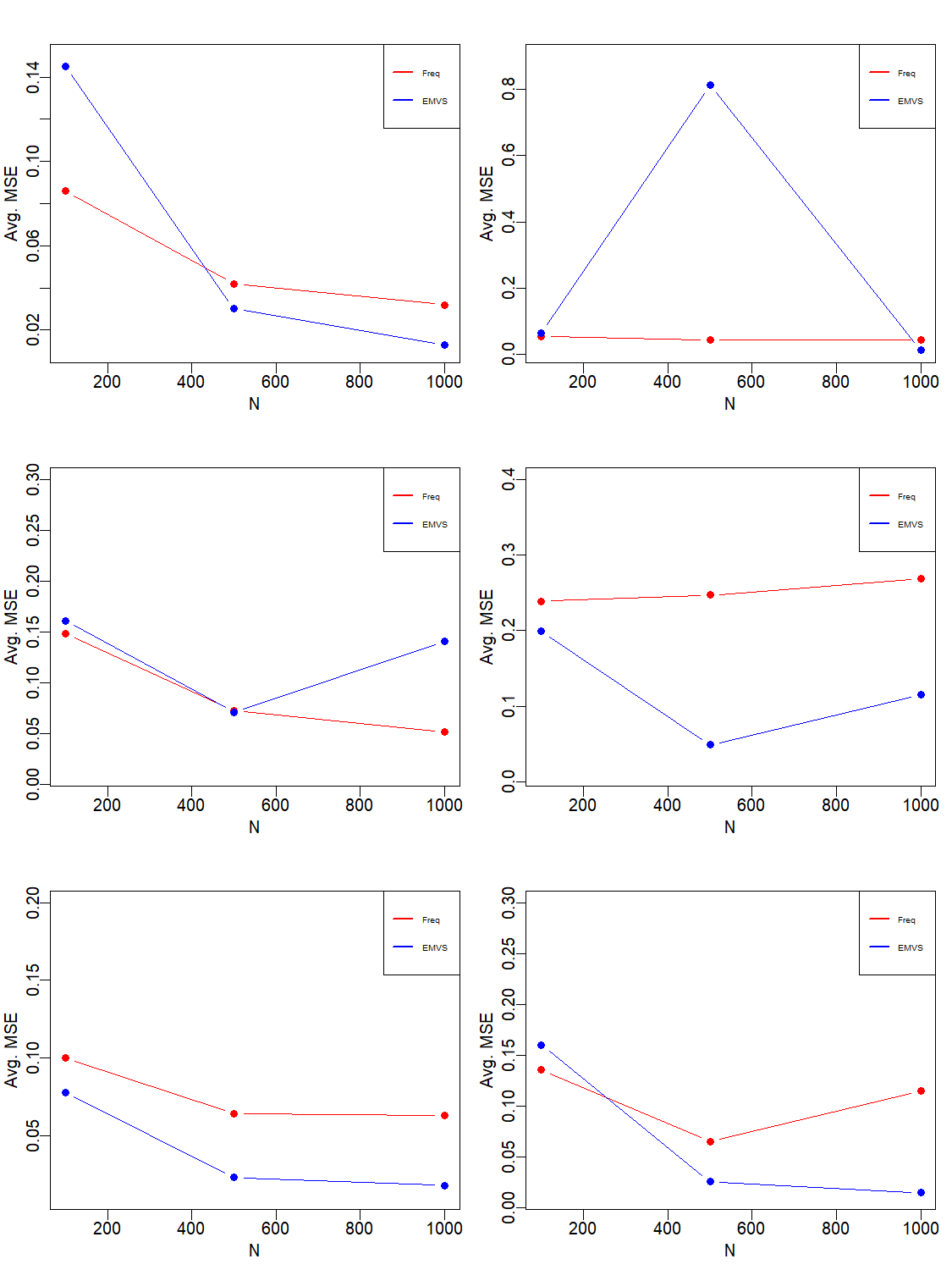}
\end{figure}

The corresponding pictures when both the arms are modeled nonparametrically are presented in Figures \ref{fig-2arm} and \ref{fig-2arm-mse}.
Tables \ref{tab2} displays the asymptotic probabilities of $d_{absorption}$ metric for  N across 100, 500, 1000 averaged across 5 replicates in quadratic setting.

\begin{table}[H] \centering 
\renewcommand\thetable{2} 
  \caption{Asymptotic probabilities of reaching the  absorbing states from the true and estimated transition functions when N=100, 500, 1000 averaged across 5 replicates in two nonparametric arm quadratic setting} 
  \label{tab2} 
\begin{tabular}{@{\extracolsep{5pt}} ccccccc} 
\\[-1.8ex]\hline 
\hline \\[-1.8ex] 
\multicolumn{1}{c}{\small N=100} &\multicolumn{3}{c}{} &
\multicolumn{3}{c}{} \\ \hline
\multicolumn{1}{c}{ } 
 &\multicolumn{3}{c}{ \small Frequentist } &
\multicolumn{3}{c}{\small Bayesian } \\
\hline
 &  \small $p$
& \small $\hat{p}$
& \small $\| p-\hat{p}\|$
& \small $p$
& \small $\hat{p}$
& \small $\| p-\hat{p}\|$ \\ 
\hline \\[-1.8ex] 
\small State 1 (censored) & \small 0.000 & \small 0.001
&  \small 0.001 & \small  0 & \small 0
& \small  0\\
\hline
\small State 2 & \small 0.622 & \small 0.599
& \small  0.023 & \small  0.601 & \small 0.592
& \small 0.009\\
\hline
\small State 3 & \small 0.377 & \small 0.400
&  \small 0.023 & \small 0.399 & \small 0.407
&  \small 0.008
\\
\hline
\hline \\[-1.8ex]
\multicolumn{1}{c}{\small N=500} &\multicolumn{3}{c}{} &
\multicolumn{3}{c}{} \\ \hline
\hline
 & \small $p$
& \small $\hat{p}$
& \small $\| p-\hat{p}\|$
& \small $p$
& \small $\hat{p}$
& \small $\| p-\hat{p}\|$
\\
\hline
\small State 1 (censored) & \small 0.001 & \small 0.001
&  \small 0 &  \small 0 & \small 0
&  \small 0\\
\hline
\small State 2 & \small 0.600 & \small 0.588
&  \small 0.012 &  \small 0.626 & \small 0.606
&  \small 0.02\\
\hline
\small State 3 & \small 0.399 & \small 0.412
&  \small 0.013 &  \small 0.374 & \small 0.393
&  \small 0.019
\\ \hline
\hline \\[-1.8ex] 
\multicolumn{1}{c}{\small N=1000} &\multicolumn{3}{c}{} &
\multicolumn{3}{c}{} \\ \hline 
\hline
 & \small $p$
& \small $\hat{p}$
& \small $\| p-\hat{p}\|$
& \small $p$
& \small $\hat{p}$
& \small $\| p-\hat{p}\|$
\\
\hline
\small State 1 (censored) & \small 0 & \small 0.001
&  \small 0.001 &  \small 0 & \small 0
&  \small 0\\
\hline
\small State 2 & \small 0.62 & \small 0.616
&  \small 0.004 &  \small 0.603 & \small 0.6
&  \small 0.003\\
\hline
\small State 3 & \small 0.379 & \small 0.383
&  \small 0.004 &  \small 0.397 & \small 0.4
&  \small 0.003
\\
\hline
\end{tabular}
\end{table}

\begin{figure}[htbp]
\centering
\caption{Plots of the the true $\lngen_{12}(\cdot), \lngen_{13}(\cdot)$ and $\lngen_{24}(\cdot)$ in quadratic, cubic and quartic forms  (black) and the corresponding estimated  versions using Bayesian approach ((posterior mode, red) and frequentist approach (blue)} \label{fig-3arm}
\includegraphics[width=0.75\textwidth]{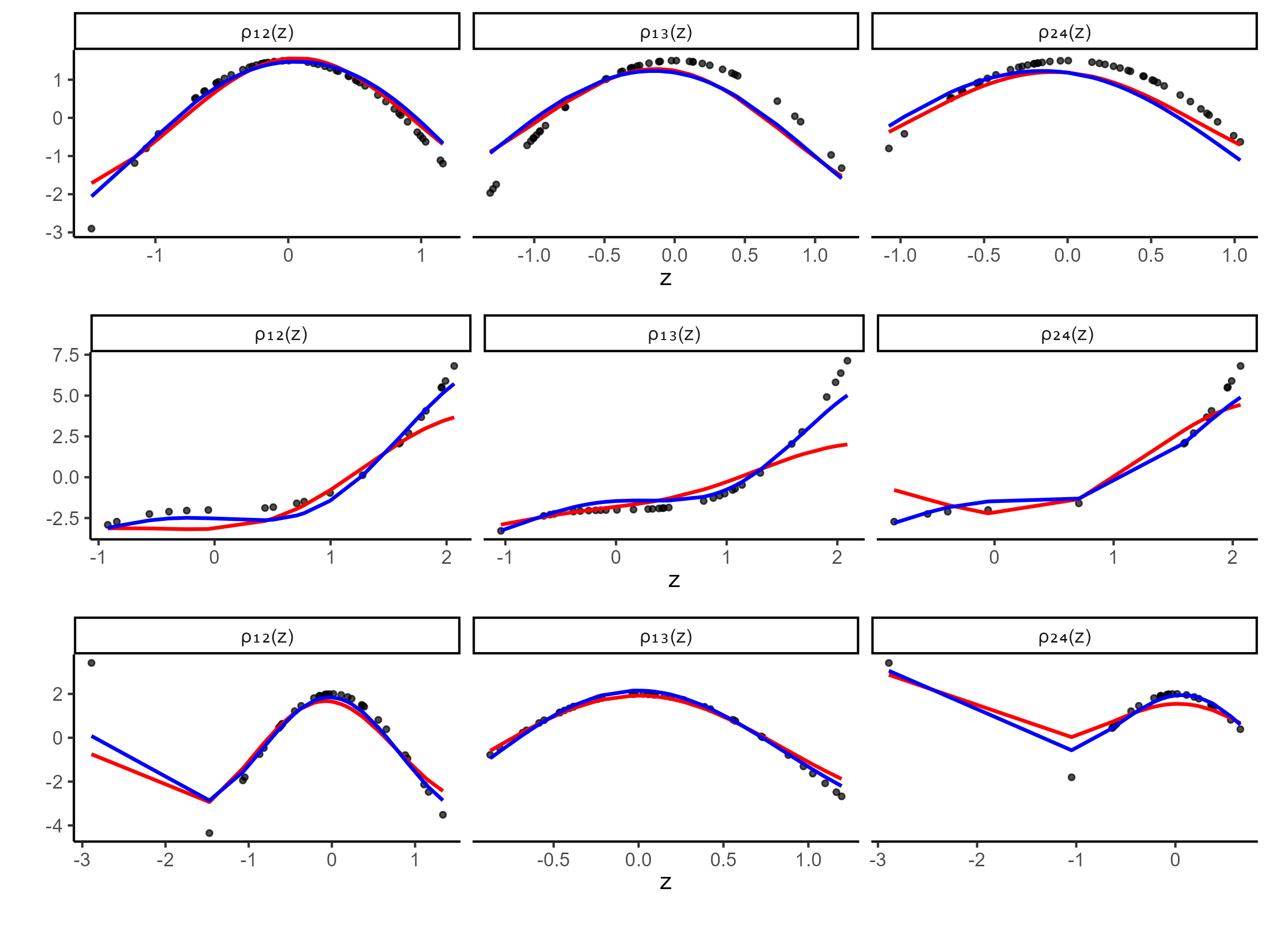}
\end{figure}
Figure \ref{fig-3arm}  compares the estimated transition functions with the true functions for the four state CTMC  where all three transition arms were modeled nonparametrically. As mentioned, we consider three cases  where the true functions  have quadratic, cubic, and quartic forms (see Table \ref{tab1}). Our results indicate no significant challenges in extending the method to multi-arm tree structures, as the general form of the likelihood remains applicable.

\section{CASE STUDY}
\label{cs}
We applied the proposed method to the  follicular cell lymphoma data \citep{pintilie2006competing} provided from the cmprskcoxmsm package by \citep{zhang2cmprskcoxmsm}. This data corresponds to a  multistate model on  three states (i.e. initial state, death and relapse) and two transition structures. 
Among  541 patients identified with early stage (I or II) follicular type lymphoma, 193 patients ($35.7 \%$) show no response to treatment  during the course of the study. Among the rest, 272 ($50.3\%$) patients relapse for the first time after treatment while 76 patients ($14\%$) transition to death without relapse. We fit a Cox regression model for the censoring process with respect to covariate of interest (Age, Haemoglobin), treating censoring as the event to verify no violation of assumption on non-informative censoring.

 We first estimate transition rate function by taking the single data point from a covariate matrix, e.g $z^*=c(61,110)$ and calculating the estimated transition rate by plugging into the kernel matrix formed by the entire covariate vector from sample, then we simulate 1000 Markov chains  and obtain the distribution of the estimated states for the intended patient.
We tabulate both distribution vectors of ending state from the estimated transition rate function with proposed method and Cox hazard model and compare them to the actual ending state from real dataset in Table \ref{tab3}. 
The results suggested with proposed method the estimated probability of our hypothetical patient whose Age (yrs) is 61 and Hgb (g/L) is 110  has about $55.2\%$ of chance to end up in Relapse and $44.7\%$ of transitioning into death state by proposed method, 
in contrast with the Cox hazard model which predicts patient transition into death rather than Relapse. While in reality, he transitioned into relapse which confirms our findings with proposed method. In order to establish the  accuracy and reliability of proposed method in predicting patient's disease transition for the entire population, we need to know the underlying censoring mechanism, which was unavailable in this dataset.

\begin{figure}[H]
\centering
\caption{Estimated transition rate functions based on Hgb in the Follicular Cell Lymphoma study:  
(a) Transition from early stage follicular lymphoma to relapse;  
(b) Transition from early stage follicular lymphoma to death.}
\label{folli}
\begin{tabular}{cc}
(a) & (b) \\
\includegraphics[width=0.45\textwidth]{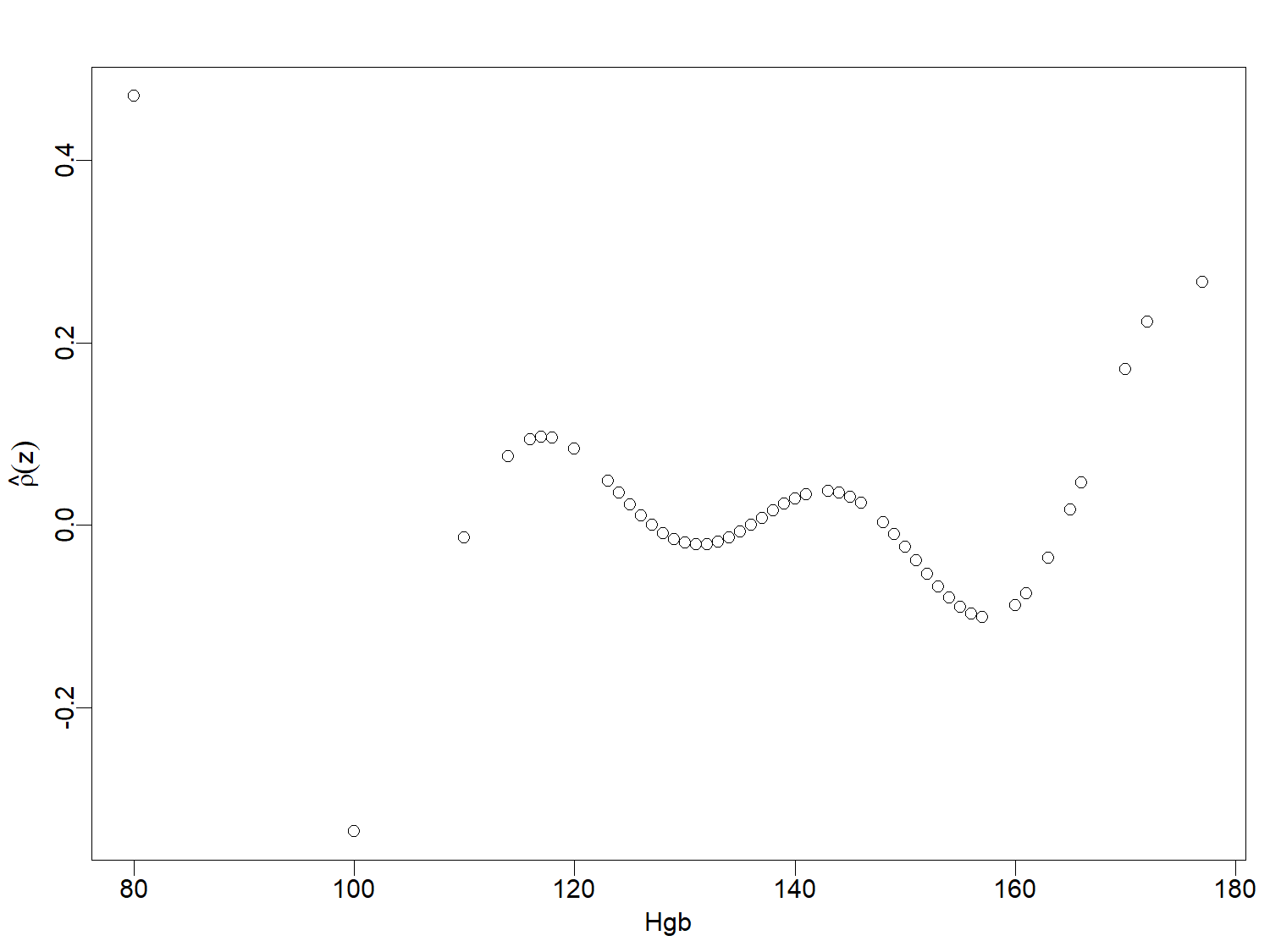} &
\includegraphics[width=0.45\textwidth]{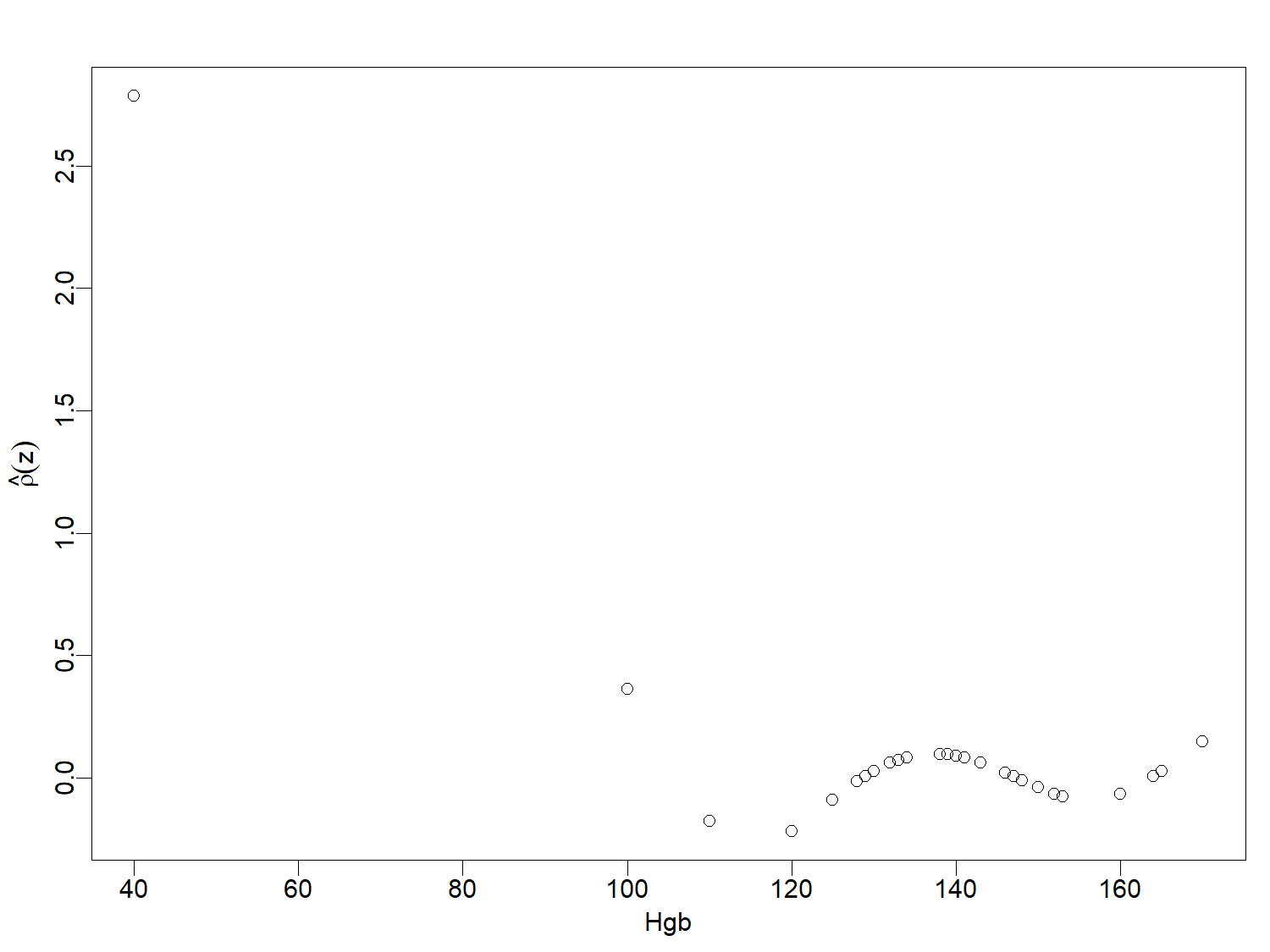} \\
\end{tabular}
\end{figure}

We plot the estimated transition rate function in Figure \ref{folli}. (a) displays the hazard rates rapidly arise but cease to increasing within interval $(100,120)$ and $(140,160)$, decreasing within $(120,140)$ then gradually increasing beyond 160 as Hgb increases in patients  transitioning from initial state to Relapse. But for patients who transition from initial state to death, Hgb versus the transition rates yields opposite trends within interval $(100,120)$. Our methods reveal the critical information about the nonlinear effect induced by a specific covariate on transition rate which is generally beyond the scope of the simple linear link function with traditional CoxPH models. 

\begin{table}[H]
\centering
\renewcommand\thetable{3} 
\caption{Asymptotic distribution of ending states averaged over 1000 Markov chains in  Follicular cell lymphoma study for patient whose Age=61 yrs, Hgb=110 g/L; $p_{\hat{\beta}}$ denotes the distribution vector produced by the proposed method, $p_{\hat{\beta}_M}$ denotes the distribution vector produced by estimates from \textit{mstate} package.} \label{tab3}
\resizebox{.4 \textwidth}{!}
{\begin{tabular}{cccccc}
\\ [-1.8ex] \hline
\hline \\[-1.8ex] 
&$p_{\hat{\beta}}$& $p_{\hat{\beta}_M}$ & $\| p_{\hat{\beta}}-p_{\hat{\beta}_M}\|$ \\
\hline
\\ [-1.8ex] 
Censored & .001 &.046&.008\\
Relapse &.552&.440 & .112\\
Death &.447 & .551 & .104\\
\hline
\end{tabular}
}
\end{table}
%

\section{DISCUSSION}
We have proposed a flexible framework for nonparametric estimation of covariate-dependent transition dynamics in CTMCs by leveraging the mathematical structure of Reproducing Kernel Hilbert Spaces (RKHS). The objective of this work is to develop a data-driven model for CTMC transition rates, avoiding the restrictive assumptions of simple parametric link functions.  Our approach relies on a generalized Representer Theorem to reduce an infinite-dimensional inference problem to a finite-dimensional optimization involving kernel expansions. This reduction enables tractable estimation of nonlinear transition functions in multistate models.

A central challenge we confront is the high dimensionality of the parameter vector inherent in RKHS representations. To address this, we impose regularization-ridge penalties in the Frequentist case and sparsity-inducing spike and slab priors in the Bayesian version. The complex structure of the likelihood function precludes any closed-form estimators or efficient Gibbs samplers. Attempts to construct Metropolis-Hastings samplers for full posterior exploration were hindered by low acceptance probabilities. To overcome this we adapt EMVS algorithm for estimating the posterior mode of  CTMC likelihoods under spike and slab priors where inclusion probabilities are modeled via Beta-Binomial priors. While EMVS does not enable full posterior sampling, it efficiently identifies sparse subsets of the coefficients of kernel-expansion and provides a viable alternative to more costly MCMC methods.

Although we focus on time-homogeneous CTMCs, the framework can easily be extended to allow time-inhomogeneous transition functions by constructing kernels over the joint space of covariates and time. This extension can be relevant for applications in systems biology and gene regulatory networks, where time-varying dynamics are common.

Finally, our framework allows for natural extensions. One practical direction involves hybrid modeling of transitions --- combining nonparametric and parametric representations across different branches. This is particularly appealing when modeling a mixture of complex biomarkers alongside simpler demographic covariates. The semiparametric design could improve computational efficiency while enhancing interpretability in applied settings.

\bibliographystyle{plainnat}

\bibliography{bib24}

\begin{thebibliography}{25}
\providecommand{\natexlab}[1]{#1}
\providecommand{\url}[1]{\texttt{#1}}
\expandafter\ifx\csname urlstyle\endcsname\relax
  \providecommand{\doi}[1]{doi: #1}\else
  \providecommand{\doi}{doi: \begingroup \urlstyle{rm}\Url}\fi

\bibitem[Aalen et~al.(2001)Aalen, Borgan, and Fekj{\ae}r]{aalen2001covariate}
Odd~O Aalen, {\O}rnulf Borgan, and Harald Fekj{\ae}r.
\newblock Covariate adjustment of event histories estimated from markov chains:
  the additive approach, 2001.

\bibitem[Carvalho et~al.(2009)Carvalho, Polson, and
  Scott]{carvalho2009handling}
Carlos~M Carvalho, Nicholas~G Polson, and James~G Scott.
\newblock Handling sparsity via the horseshoe.
\newblock In \emph{Artificial intelligence and statistics}, pages 73--80. PMLR,
  2009.

\bibitem[Carvalho et~al.(2010)Carvalho, Polson, and
  Scott]{carvalho2010horseshoe}
Carlos~M Carvalho, Nicholas~G Polson, and James~G Scott.
\newblock The horseshoe estimator for sparse signals.
\newblock \emph{Biometrika}, 97\penalty0 (2):\penalty0 465--480, 2010.

\bibitem[Cox(1972)]{cox1972regression}
David~R Cox.
\newblock Regression models and life-tables.
\newblock \emph{Journal of the Royal Statistical Society: Series B
  (Methodological)}, 34\penalty0 (2):\penalty0 187--202, 1972.

\bibitem[Cremaschi et~al.(2022)Cremaschi, Argiento, De~Iorio, Shirong, Chong,
  Meaney, and Kee]{crem}
Andrea Cremaschi, Raffaele Argiento, Maria De~Iorio, Cai Shirong, Yap~Seng
  Chong, Michael Meaney, and Michelle Kee.
\newblock Seemingly unrelated multi-state processes: a bayesian semiparametric
  approach.
\newblock \emph{Bayesian Analysis}, 1\penalty0 (1):\penalty0 1--23, 2022.

\bibitem[De~Wreede et~al.(2010)De~Wreede, Fiocco, and Putter]{de2010mstate}
Liesbeth~C De~Wreede, Marta Fiocco, and Hein Putter.
\newblock The mstate package for estimation and prediction in non-and
  semi-parametric multi-state and competing risks models.
\newblock \emph{Computer methods and programs in biomedicine}, 99\penalty0
  (3):\penalty0 261--274, 2010.

\bibitem[Fiocco et~al.(2008)Fiocco, Putter, and van Houwelingen]{fico}
Marta Fiocco, Hein Putter, and Hans~C van Houwelingen.
\newblock Reduced-rank proportional hazards regression and simulation-based
  prediction for multi-state models.
\newblock \emph{Statistics in Medicine}, 27\penalty0 (21):\penalty0 4340--4358,
  2008.

\bibitem[Ganguly et~al.(2023)Ganguly, Mitra, and Zhou]{ganguly2023infinite}
Arnab Ganguly, Riten Mitra, and Jinpu Zhou.
\newblock Infinite-dimensional optimization and bayesian nonparametric learning
  of stochastic differential equations.
\newblock \emph{Journal of Machine Learning Research}, 24\penalty0
  (159):\penalty0 1--39, 2023.

\bibitem[George and McCulloch(1997)]{GM97}
Edward~I George and Robert~E McCulloch.
\newblock Approaches for bayesian variable selection.
\newblock \emph{Statistica sinica}, pages 339--373, 1997.

\bibitem[Hatami et~al.(2024)Hatami, Ocampo, Graham, Nichols, and
  Ganjgahi]{hatami2024scalable}
Farhad Hatami, Alex Ocampo, Gordon Graham, Thomas~E Nichols, and Habib
  Ganjgahi.
\newblock A scalable approach for continuous time markov models with
  covariates.
\newblock \emph{Biostatistics}, 25\penalty0 (3):\penalty0 681--701, 2024.

\bibitem[Kalbfleisch and Lawless(1985)]{kalbfleisch1985analysis}
JD~Kalbfleisch and Jerald~Franklin Lawless.
\newblock The analysis of panel data under a markov assumption.
\newblock \emph{Journal of the american statistical association}, 80\penalty0
  (392):\penalty0 863--871, 1985.

\bibitem[Kalbfleisch and Lawless(1988)]{kalbfleisch1988likelihood}
JD~Kalbfleisch and JF~Lawless.
\newblock Likelihood analysis of multi-state models for disease incidence and
  mortality.
\newblock \emph{Statistics in medicine}, 7\penalty0 (1-2):\penalty0 149--160,
  1988.

\bibitem[Kaur et~al.(2022)Kaur, Geistkemper, Mitra, and Becker]{kaur22}
Ramandeep Kaur, Anne Geistkemper, Riten Mitra, and Ellen Becker.
\newblock Impact of respiratory therapist education on discharge outcomes of
  subjects with covid-19, 2022.

\bibitem[Kneib and Hennerfeind(2008)]{kneib2008bayesian}
Thomas Kneib and Andrea Hennerfeind.
\newblock Bayesian semi parametric multi-state models.
\newblock \emph{Statistical Modelling}, 8\penalty0 (2):\penalty0 169--198,
  2008.

\bibitem[Le-Rademacher et~al.(2018)Le-Rademacher, Peterson, Therneau, Sanford,
  Stone, and Mandrekar]{le2018application}
Jennifer~G Le-Rademacher, Ryan~A Peterson, Terry~M Therneau, Ben~L Sanford,
  Richard~M Stone, and Sumithra~J Mandrekar.
\newblock Application of multi-state models in cancer clinical trials.
\newblock \emph{Clinical Trials}, 15\penalty0 (5):\penalty0 489--498, 2018.

\bibitem[Marshall and Jones(1995)]{marshall1995multi}
Guillermo Marshall and Richard~H Jones.
\newblock Multi-state models and diabetic retinopathy.
\newblock \emph{Statistics in medicine}, 14\penalty0 (18):\penalty0 1975--1983,
  1995.

\bibitem[Meira-Machado et~al.(2009)Meira-Machado, de~Una-{\'A}lvarez,
  Cadarso-Su{\'a}rez, and Andersen]{meira2009multi}
Lu{\'\i}s Meira-Machado, Jacobo de~Una-{\'A}lvarez, Carmen Cadarso-Su{\'a}rez,
  and Per~K Andersen.
\newblock Multi-state models for the analysis of time-to-event data.
\newblock \emph{Statistical methods in medical research}, 18\penalty0
  (2):\penalty0 195--222, 2009.

\bibitem[Park and Casella(2008)]{park2008bayesian}
Trevor Park and George Casella.
\newblock The bayesian lasso.
\newblock \emph{Journal of the American Statistical Association}, 103\penalty0
  (482):\penalty0 681--686, 2008.

\bibitem[Pintilie(2006)]{pintilie2006competing}
Melania Pintilie.
\newblock \emph{Competing risks: a practical perspective}.
\newblock John Wiley \& Sons, 2006.

\bibitem[Putter et~al.(2007)Putter, Fiocco, and Geskus]{putter2007tutorial}
Hein Putter, Marta Fiocco, and Ronald~B Geskus.
\newblock Tutorial in biostatistics: competing risks and multi-state models.
\newblock \emph{Statistics in medicine}, 26\penalty0 (11):\penalty0 2389--2430,
  2007.

\bibitem[Ro{\v{c}}kov{\'a} and George(2014)]{rovckova2014emvs}
Veronika Ro{\v{c}}kov{\'a} and Edward~I George.
\newblock Emvs: The em approach to bayesian variable selection.
\newblock \emph{Journal of the American Statistical Association}, 109\penalty0
  (506):\penalty0 828--846, 2014.

\bibitem[Tipping(2001)]{tipp}
Michael~E. Tipping.
\newblock Sparse {B}ayesian learning and the relevance vector machine.
\newblock \emph{J. Mach. Learn. Res.}, 1\penalty0 (3):\penalty0 211--244, 2001.
\newblock ISSN 1532-4435.

\bibitem[van Erp et~al.(2019)van Erp, Oberski, and Mulder]{VOM19}
Sara van Erp, Daniel~L. Oberski, and Joris Mulder.
\newblock Shrinkage priors for {B}ayesian penalized regression.
\newblock \emph{J. Math. Psych.}, 89:\penalty0 31--50, 2019.
\newblock ISSN 0022-2496.

\bibitem[Zhang and Xu(2021)]{zhang2cmprskcoxmsm}
Yiran Zhang and Ronghui Xu.
\newblock \emph{cmprskcoxmsm: Use IPW to Estimate Treatment Effect under
  Competing Risks}, 2021.
\newblock URL \url{https://CRAN.R-project.org/package=cmprskcoxmsm}.
\newblock R package version 0.2.1.

\bibitem[Zhao et~al.(2016)Zhao, Wang, Cumberworth, Gsponer, de~Freitas, and
  Bouchard-C{\^o}t{\'e}]{zhao2016bayesian}
Tingting Zhao, Ziyu Wang, Alexander Cumberworth, Joerg Gsponer, Nando
  de~Freitas, and Alexandre Bouchard-C{\^o}t{\'e}.
\newblock Bayesian analysis of continuous time markov chains with application
  to phylogenetic modelling.
\newblock 2016.

\end{thebibliography}
\end{document}